\documentclass[conference]{IEEEtran}
\IEEEoverridecommandlockouts

\usepackage{cite}
\usepackage{url}
\usepackage{amsmath,amssymb,amsfonts}
\usepackage{algorithmic}
\usepackage{graphicx}
\usepackage{textcomp}
\usepackage{xcolor}
\def\BibTeX{{\rm B\kern-.05em{\sc i\kern-.025em b}\kern-.08em
    T\kern-.1667em\lower.7ex\hbox{E}\kern-.125emX}}
\begin{document}

\title{Real-Time and Accurate: Zero-shot High-Fidelity Singing Voice Conversion with Multi-Condition Flow Synthesis\\
}

\author{\IEEEauthorblockN{1\textsuperscript{st} Hui Li}
\IEEEauthorblockA{\textit{Harbin Engineering University}\\
Harbin, China \\
lalala.lh1313@gmail.com}
\and
\IEEEauthorblockN{2\textsuperscript{nd} Hongyu Wang}
\IEEEauthorblockA{\textit{Harbin Engineering University}\\
Harbin, China \\
hongyuwang0414@gmail.com}
\and
\IEEEauthorblockN{3\textsuperscript{rd} Zhijin Chen}
\IEEEauthorblockA{\textit{Harbin Engineering University}\\
Harbin, China \\
18731572303@163.com}
\and
\IEEEauthorblockN{4\textsuperscript{th} Bohan Sun}
\IEEEauthorblockA{\textit{Harbin Engineering University}\\
Harbin, China \\
bohansun0531@163.com}
\and
\IEEEauthorblockN{5\textsuperscript{th} Bo Li}
\IEEEauthorblockA{\textit{Harbin Engineering University}\\
Harbin, China \\
libo520209@126.com}
}

\maketitle

\begin{abstract}
Singing voice conversion is to convert the source singing voice into the target singing voice except for the content. Currently, flow-based models can complete the task of voice conversion, but they struggle to effectively extract latent variables in the more rhythmically rich and emotionally expressive task of singing voice conversion, while also facing issues with low efficiency in speech processing. In this paper, we propose a high-fidelity flow-based model based on multi-decoupling feature constraints called RASVC, which enhances the capture of vocal details by integrating multiple latent attribute encoders. We also use  Multi-stream inverse short-time Fourier transform(MS-iSTFT) to enhance the speed of speech processing by skipping some complicated decoder processing steps. We compare the synthesized singing voice with other models from multiple dimensions, and our proposed model is highly consistent with the current state-of-the-art, with the demo which is available at \url{https://lazycat1119.github.io/RASVC-demo/}
\end{abstract}


\section{Introduction}
The main task of singing voice conversion (SVC) is to convert the voice of the source singer into the voice of the target singer, while retaining the original content and melody. High-quality singing conversion not only needs to accurately retain the content information but also needs to capture and reconstruct the information of the singer's musical elements (including timbre, pitch, rhythm, emotion, etc.)[1][2]
Speech content should fully contain all the content information. The quality of PPGS-based representation [3][4][5]proposed by early researchers depends heavily on the performance of the ASR system, which requires a lot of labeled data for training. Later, SSL-based representation[6]can decouple the speaker's information, but its generalization ability and stability are poor. VAE-based content embedding is stable and can capture deep information at the same time[7][8], but it still entangles speaker information and 
only approximately optimizes the evidence lower bound rather than directly calculating the exact log-likelihood, such as WavLM[9] and ContentVec[10][11]. These models all take continuous variables as the existing form of content information, and now the proposed speech discretization accelerates the speed of speech processing, such as VQ-VAE[12][13] and RVQ[14], but they are at the expense of losing some content information.

On the generation framework, one kind of model adopts an end-to-end system, such as FastSpeech2[15], AutoVC[16], or Transformer TTS[17][18], and the other kind of model generates acoustic features (such as Mel spectrogram). It uses a vocoder to generate speech waveforms. At present, the vocoder based on GANs is popular, and high-fidelity waveforms, such as HiFi-GAN[19] and MelGAN[20], are generated through confrontation training, but the problem is that the decoding speed is too slow. In recent years, the vocoder model based on Flow has also been widely concerned[21][22]. It is a generation model that makes the optimization goal solvable by constructing reversible transformations, such as Nice[23], RealNVP[24], Glow[25], and WaveGlow[26]. They accelerate the process of speech generation through parallelization. However, in these speech models, the flow input conditions are few, and only the speaker's identity embedding and content embedding are considered, which leads to the lack of naturalness and expressiveness of the generated speech and makes it difficult to perform the task of song conversion.

To solve the above problems, we use the HuBERT-Soft model[27] to complete content extraction, which not only reduces the dependence on a large number of labeled data but also directly finds an accurate balance between discrete features and continuous features and separates content information. We also use the flow model to directly maximize the exact log-likelihood, reversibly transforming simple distributions into complex ones. Additionally, we introduce timbre encoder, pitch encoder, and emotion encoder as conditions, which improves the information integrity of the generated speech. Finally, we choose MS-iSTFT to skip some complicated decoder processing steps, speed up the processing of the decoder module.

In this study, we have realized a fast and high-quality singing conversion model. The main contributions of this paper can be summarized as follows:

\begin{figure*}[ht]
    \centering
    \includegraphics[width=0.8\linewidth]{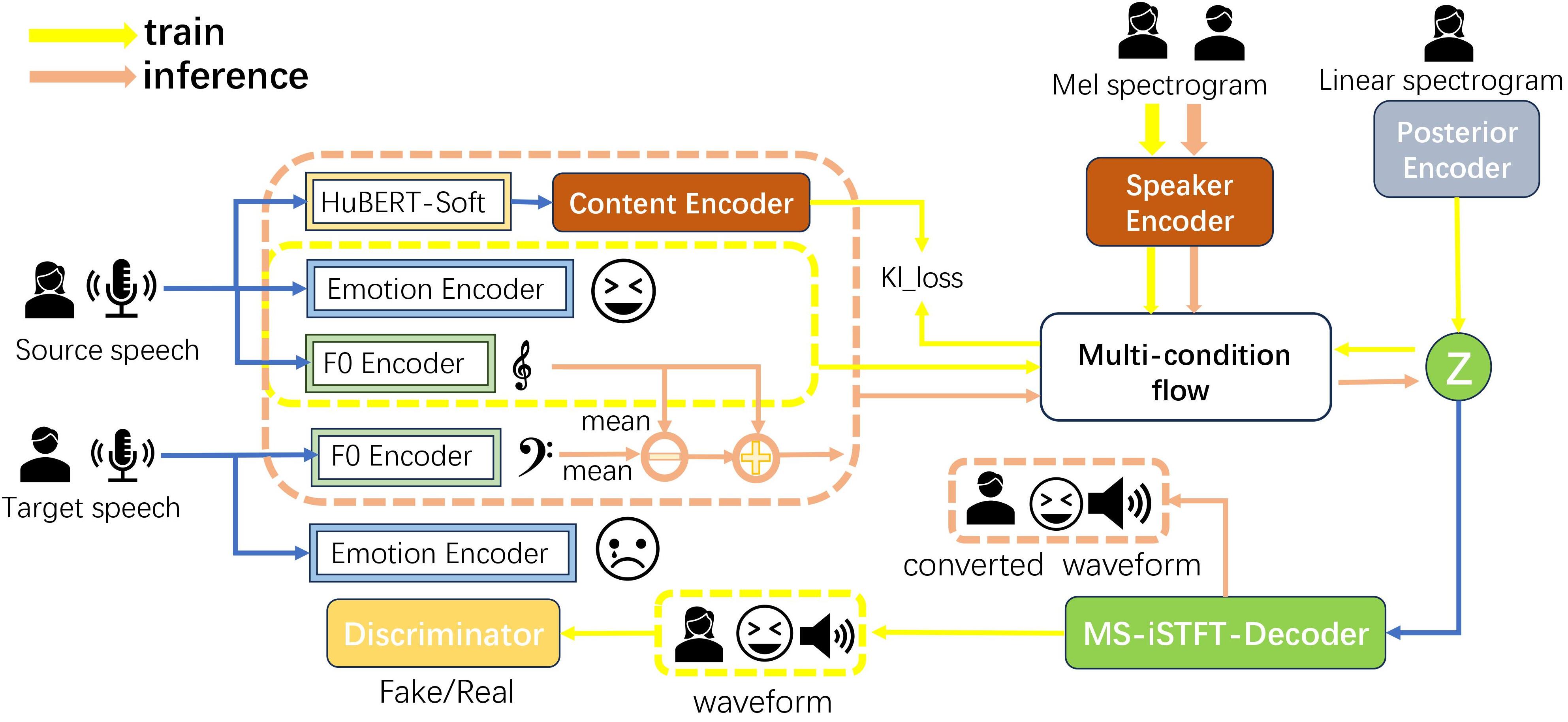}
    \caption{Overview of model structure}
    \label{fig:1}
\end{figure*}

1. We propose to use the HuBERT-Soft model to gain soft speech units by predicting a distribution over the discrete units, which 
effectively extract the content information in singing voice conversion.

2. We propose a Multi-condition-based Flow, which not only extracts the speaker's timbre but also introduces pitch and emotion as the condition of the flow by extra extractors, which greatly improves the naturalness and expressiveness of singing voice conversion.

3. We propose to replace some complicated processing steps of the decoder in VITS with Multi-stream inverse short-time Fourier transform to directly convert from frequency domain features to time domain waveforms, which greatly enhances the speed of synthesis.

\section{Method}
\subsection{Overall pipeline}
As shown in Figure 1, our model is based on VITS model[28]. Our model includes an F0 encoder, an emotion encoder, a speaker encoder, a content encoder, a Multi-condition flow, an MS-iSTFT-Decoder, and a discriminator. In the following sections, we focus on describing the aforementioned multiple encoders, the Multi-condition flow, the MS-iSTFT-Decoder, the loss function we used and the strategy in training and inference.

\subsection{Multi-condition encoder}
In this section, we will introduce four encoders used to assist speech generation when entering the flow-based model, namely content Encoder, speaker encoder, f0 encoder and emotion encoder.
\subsubsection{Content Encoder}
Content encoding includes the Hubert-soft model and a content encoder. A major challenge in self-supervised speech representation learning is that speech contains multiple units and there are typically no discrete words or characters as input. By inputting the source speech into the pre-trained Hubert-soft model, we learn the latent distribution of the discrete variables of speech, which results in an aligned sequence \( h(x) = Z = [z_1, \ldots, z_T] \), where each unit has 1024-dimensional features. These features are then fed into the content encoder and transformed into a lower-dimensional content embedding \( Z_c \).
\subsubsection{Speaker Encoder}
We input the Mel-spectrograms feature of the target singer into a pre-trained LSTM framework to obtain embeddings with identity discrimination.
\subsubsection{F0 Encoder}
We employ a monophonic pitch tracker based on a deep convolutional neural network to obtain the continuous fundamental frequency (F0). Considering the  differences in fundamental frequencies between male and female voices can lead to unnatural-sounding singing, during the inference phase, we add the average difference in fundamental frequencies between the target and source voices to the source singing's F0. This results in a final F0 feature that not only preserves the dynamic characteristics of the original song's vocal track but also retains the vocal traits of the target speaker.
\subsubsection{Emotion Encoder}
We use a self-supervised and pre-trained universal emotional expression model, emotion2vec[29], to extract the emotional features of each song, serving as a constraint for the flow model, which is beneficial for a more natural transition in song conversion and the embodiment of the song's artistic conception.

\subsection{Flow with multiple decoupling attribute constraints}
We apply the normalizing flow composed of multiple coupled affine layers to the decomposed simple prior distribution. Through the reversible transformation of the simple prior distribution, it directly maxes the exact log-likelihood and is transformed into a more complex distribution. At the same time, we use the decoupling characteristics of multiple encoders as constraints of normalizing flow, which improves the expressive ability of prior distribution and can capture the distribution characteristics of real samples better. We use $f_{\theta}$ to represent the standardized flow. According to the variable transformation theorem, the prior distribution can be rewritten as:
\begin{equation}
p_{\theta}(z|c) = \mathcal{N}(f_{\theta}(z); \mu_{\theta}(c), \sigma_{\theta}(c)) \left|\det \frac{\partial f_{\theta}(z)}{\partial z}\right|
\end{equation}

where $c$ here is the output embedding of the priori encoder. $c$ can be expressed as follows:

\begin{equation}
c=[C_{content},C_{speaker},C_{emotion},C_{f0}]
\end{equation}

\subsection{MS-iSTFT-Decoder}
We utilize iSTFT to replace certain repetitive network layers in the previous HiFi-GAN vocoder by introducing the computation of phase and amplitude, converting latent embedding into continuous time-domain waveforms. This approach effectively reduces computational load and accelerates the audio synthesis process.

\subsection{Multi-loss construction}
Similar to VITS, we integrate Variational Autoencoder (VAE) and Generative Adversarial Network (GAN) methodologies into our training process. The overall loss is articulated as follows:

\begin{equation}
L_{\text{total}} = L_{\text{recon}} + L_{\text{kl}} + L_{\text{adv}}(G) + L_{\text{fm}}(G)
\end{equation}

We transform the generated waveform to the mel-spectrogram domain and calculate the \( L_1 \) loss against the source speech's mel-spectrogram to serve as the reconstruction loss:

\begin{equation}
L_{\text{recon}} = \left| X_{\text{mel}} - \hat{X}_{\text{mel} }\right|
\end{equation}

KL loss is used to narrow the gap between the prior encoder and the posterior encoder in terms of their distributions. The formula for KL loss is as follows:

\begin{equation}
L_{\text{kl}} = \log q_{\phi}(z|x_{\text{lin}}) - \log p_{\theta}(z|c)
\end{equation}
Among which, $x_{\text{lin}}$ represents the linear-scale spectrogram of the speech and the distribution of the posterior encoder can be expressed as:\begin{equation}
q_{\phi}(z|x_{\text{lin}}) = \mathcal{N}(z; \mu_{\phi}(x_{\text{lin}}), \sigma_{\phi}(x_{\text{lin}}))
\end{equation}

By introducing a pre-trained discriminator \( D \), we verify the authenticity of the speech generated by the decoder \( G \), leveraging \( L_{\text{adv}}(G) \) for supervision. Furthermore, we incorporate an additional feature-matching Loss \( L_{\text{fm}}(G) \) to ensure consistency in the reconstruction loss measured within the discriminator's hidden layers between the generated and authentic speech.

\subsection{Training and inference strategy}
During the training process, we input the same piece of speech into the network. The linear-scale spectrogram of the speech is sent to the Posterior Encoder to obtain the latent variable \( Z \), which is then fed into the Decoder to produce synthesized speech. We calculate \( L_{\text{recon}} \),  \( L_{\text{adv}}(G) \), and \( L_{\text{fm}}(G) \). Simultaneously, under the constraints of multiple disentangled attributes, the latent variable \( Z \) is transformed through a flow to \( Z_p \), and \( L_{\text{kl}} \) is computed with the content embedding derived from the Content Encoder.

During the inference process, we input the speech from two different speakers into the network. The content and emotion originate from the source speech, the timbre comes from the target speech, and the pitch is derived from a combined calculation of both the source and target speech. These elements serve as constraints for the flow, resulting in the latent variable 
\( Z \), which is then fed into the decoder to produce synthesized speech.

\section{Experiment and Results}
\subsection{Datasets}
We conduct experiments on four datasets: VCTK, Opensinger, M4singer and NUS-48E. We take the model weights that have been trained on the VCTK dataset (which contains 44 hours of speeches from 107 English speakers with various accents) as the starting point and then continue to train on the Opensinger dataset (a large-scale Chinese singing voice dataset) to learn the distinctive characteristics of Chinese singing voices. Finally, we use the M4singer dataset (a multi-style and multi-singer Chinese singing voice dataset) and the NUS-48E (an English singing voice dataset) to test the evaluation model's capabilities in zero-shot singing voice conversion (SVC) and cross-domain conversion.
\subsection{Detail}
In the training stage, the speech resampling frequency is 16khz, and the speech Mel-spectrogram is extracted by 512-point fast Fourier transform and 512-point window calculation. The model is preloaded with the pre-training weights of Quickvc[30] (trained on VCTK for 2 weeks), then trained on the Opensinger dataset for three days, and tested on M4singer and NUS-48E. At the same time, other experimental schemes such as FastSVC[31] and SoVITS-Flow are compared. All models have been fully iteratively trained on a single NVIDIA 3090ti, with a batch size of 64 and learning rates of 1e-4 and 5e-5 respectively.
\subsection{Results and analysis}

\begin{table*}[htbp]
\centering
\caption{Model comparison}
\label{table}
\small
\setlength{\tabcolsep}{5pt}
\begin{tabular}{|c|c|c|c|c|c|}
\hline
\textbf{Method} & \textbf{MOS/Naturalness} & \textbf{MOS/Similarity}  & \textbf{PESQ} & \textbf{Voice similarity} & \textbf{RTF(GPU)}\\
\hline
FastSVC &  3.52 ± 0.10 &  3.27 ± 0.22 & - & 0.433& 0.031 \\
SoVITS-Flow & 3.10 ± 0.22 &  2.90 ± 0.23  & 2.486 & 0.585 & 0.008\\
CoMoSVC & $\mathbf{4.27 \pm 0.16}$ & 4.00 ± 0.19  & $\mathbf{2.948}$ & 0.585 & $\mathbf{0.006}$\\
DiffSVC & 4.23 ± 0.19 &  3.95 ± 0.21  & 2.917 &  0.598& 0.278\\
Multi-F0 Model &3.85 ± 0.02 & 3.47 ± 0.04  & - & - & - \\
UCD-SVC  & 3.06 ± 0.08 & 2.67 ± 0.18  & - & 0.588 & 0.103 \\
\hline
$\mathbf{Ours}$ &4.14 ± 0.17 $\downarrow$ & $\mathbf{4.02 \pm 0.19}$$\uparrow$ & 2.834 $\downarrow$& $\mathbf{0.603}$ $\uparrow$ & 0.048 $\downarrow$\\
\hline
\multicolumn{6}{c}{Compared with the indexes of the restored songs at 16kHz, many experiments have been done} \\
\multicolumn{6}{c}{to get the average value, and bold indicates the best result.} \\
\end{tabular}
\label{tab1}
\end{table*}
\renewcommand{\thefootnote}{\alph{footnote}}
\subsubsection{Model evaluation}
We adopt MOS/Naturalness, MOS/Similarity, Pesq, Voice Similarity and RTF as the evaluation indexes of the model, such as FastSVC, SoVITS-Flow\footnote{\url{https://github.com/svc-develop-team/so-vits-svc?tab=readme-ov-fle#sovits-model}}, CoMoSVC[32], DiffSVC[33], Multi-F0 Model[34], UCD-SVC and the evaluation objects are the song conversion of different models in the dataset M4Singer → NUS-48E and the cross-language song conversion.
Table 1 shows that: (1) we evaluate the naturalness and similarity of voice by collecting people's scores on MOS values extensively, and the results show that the naturalness of our proposed RASVC model reaches 4.14, which exceeds all baseline models except CoMoSVC, which proves that the multi-conditional strategy we joined has not reduced the naturalness, but made the emotion and pitch of singing more accurate and rich; (2) in terms of MOS similarity, people have scored more than the target timbre and the converted timbre. (3) Compared with other models, the PESQ index of our proposed model has declined, which may be due to the uneven distribution of variables under most conditions, but it still remains at a high level. (4) We use a general speaker verification (ASV) model, such as ECAPA-TDNN and Comformer to evaluate the timbre of the target song and the timbre of the converted song. The results show that our model maintains the highest sound similarity in cross-domain and cross-language singing conversion, which is closely related to F0(fundamental frequency) assistance. (5) Finally, we test the model RTF (the time required for the model to process one second of speech) in a single NVIDIA 3090ti. The other models here are from their source paper values, which may be because they use better hardware devices than us.

\begin{figure}[ht]
    \centering
    \includegraphics[width=0.65\linewidth]{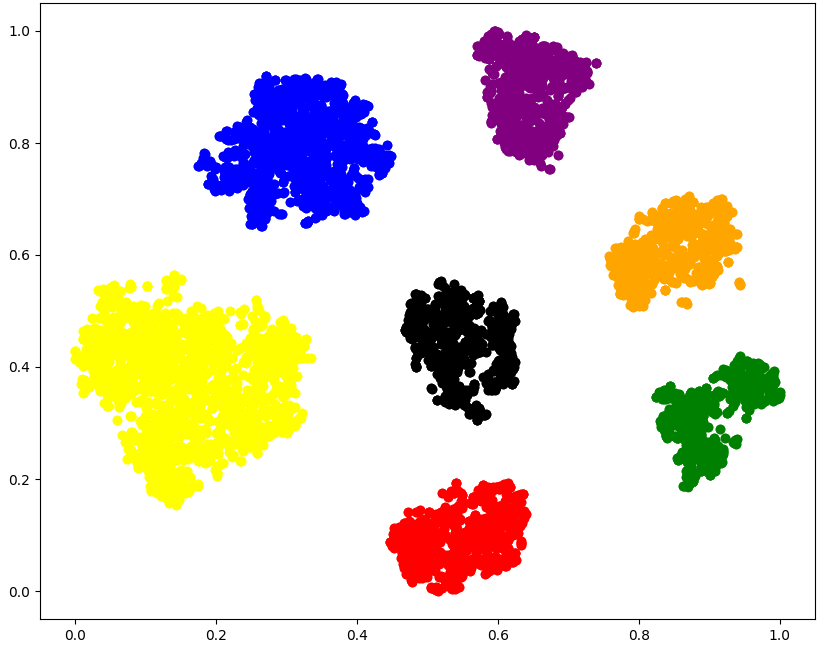}
    \caption{t-SNE visualization of converted songs}
    \label{fig:1}
\end{figure}

\subsubsection{t-SNE visualization of converted songs}
As shown in Figure 2, we will transform the songs during the training process through the speaker encoder and draw them embedded in 2D space with t-SNE. It is found that different songs are still separated after conversion, which shows that our multi-conditional embedding does not destroy the identity information of speakers, but also effectively enhances the similarity, opens the distance information between different speakers and learns reasonable representation information.

\subsubsection{Visualization analysis of Spectrogram}

As can be seen from Figure 3, when the male turns to the female, the resonance peak moves upward, and the interval of vertical stripes is obviously widened; When a woman turns to a man, the resonance peak moves downward, and the interval of vertical stripes is obviously narrowed. At the same time, the main spectral features of the original voice (such as the overall energy distribution of the spectrogram) are preserved after the two transformations. It can be seen that our model not only accurately adjusts the fundamental frequency (f0) in the transformation of male and female singing, but also effectively preserves the content of singing.

\begin{figure}[ht]
    \centering
    \includegraphics[width=1\linewidth]{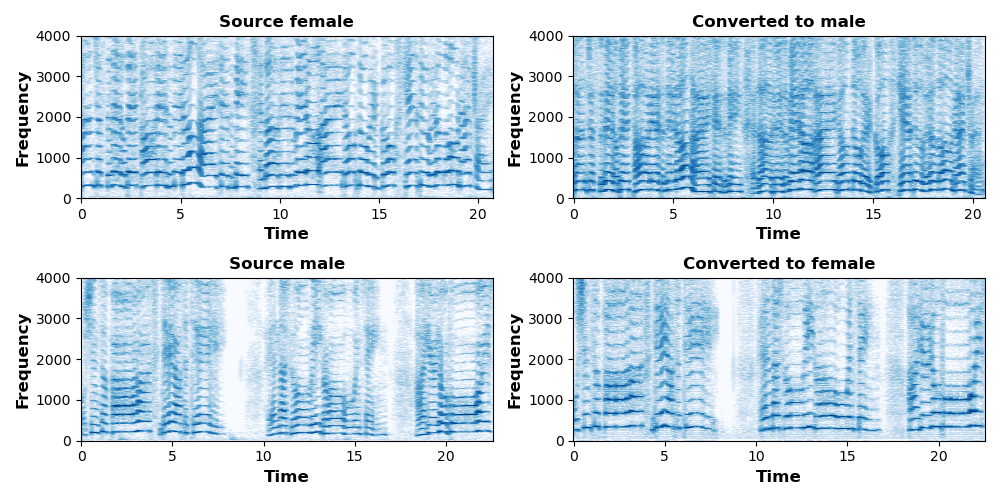}
    \caption{Visualization analysis of Spectrogram}
    \label{fig:1}
\end{figure}

\section{Conclusion}

In this paper, we propose a high-fidelity flow-based model based on multi-decoupling feature constraints. This model uses timbre, pitch, content, and emotion to assist the flow model in completing the song synthesis and conversion, and it is also used in diversified singing scenes. The inverse Fourier transform is also applied to the decoder to improve the conversion efficiency. The experimental results show that the naturalness and similarity of the speech after the conversion of our proposed model are 4.14 and 4.02. Finally, we have given a demo and will release a Pytorch trainer for singing voice conversion to promote further research in this field.

\end{document}